\documentstyle[preprint,aps]{revtex}
\begin{document}
\draft

\title{Unified Description and Canonical Reduction to Dirac's Observables of
the Four Interactions}

\author{Luca Lusanna}
\address{Sezione INFN di Firenze\\
L.go E.Fermi 2 (Arcetri)\\
50125 Firenze, Italy\\
email: lusanna@fi.infn.it}

\maketitle

Talk given at the International Workshop ``New Non Perturbative Methods and
Quantization on the Light Cone", Les Houches February 24 - March 7, 1997.

\section{Systems with Constraints.}

The standard $SU(3)\times SU(2)\times U(1)$ model of elementary particles, all
its extensions with or without supersymmatry, all variants of string theory, all
formulations of general relativity are described by singular Lagrangians.
Therefore, their Hamiltonian formulation needs Dirac's theory of 1st and 2nd
class constraints\cite{dirac} determining the submanifold of phase space 
relevant for dynamics: this means that the basic mathematical structure behind
our description of the four interactions is presymplectic geometry
(namely the theory of submanifolds of phase space with a closed degenerate
2-form; strictly speaking only 1st class constraints are associated with
presymplectic manifolds: the 2nd class ones complicate the structure). For a 
system with 1st and 2nd class constraints the physical description becomes clear
in coordinates adapted to the presymplectic submanifold. Locally in phase space,
an adapted Darboux chart can always be found by means of Shanmugadhasan's
canonical transformations \cite{sha} [strictly speaking their existence is 
proved only for finite-dimensional systems, but they underlie the existence of 
the Faddeev-Popov measure for the path integral]. The new canonical basis has: 
i) as many new momenta as 1st class constraints (Abelianization of 1st class
constraints); ii) their conjugate canonical variables (Abelianized gauge 
variables); iii) as many pairs of canonical variables as pairs of 2nd class
constraints (standard form, adapted to the chosen Abelianization, of the
irrelevant variables); iv) pairs of canonically conjugate Dirac's observables 
(canonical basis of physical variables adapted to the chosen Abelianization;
they give a trivialization of the BRST construction of observables). Putting 
equal to zero the Abelianized gauge variables one defines a local gauge of the
model. If a system with constraints admits one (or more) global
Shanmugadhasan canonical transformations, one obtains one (or more) privileged 
global gauges in which the physical Dirac observables are globally defined and
globally separated from the gauge and the irrelevant degrees of freedom [for
systems with a compact configuration space this is impossible]. These
privileged gauges (when they exist) can be called generalized Coulomb gauges.

To find them the main problem is to discover how the original canonical
variables depend upon the gauge variables. This can be achieved by solving (if
possible) the so-called multitemporal equations
[see Refs.\cite{re}a), b)]: by considering each 
of the original 1st class constraints as the Hamiltonian for the evolution in 
a suitable parameter (called a generalized time), these equations are the 
associated Hamilton equations. If one succeeds in solving these equations
(which are formally integrable due the 1st class nature of the constraints),
one finds the the suitable parameters are just the Abelianized gauge variables
and, then, one can construct the conjugate Abelianized 1st class constraints
and the standard form of the 2nd class ones. Every set of suitable generalized 
times gives rise to a different generalized Coulomb gauge. Let us remark that 
in certain cases it is possible to find some special global Shanmugadhasan
canonical transformations such that the effective new Hamiltonian of the system
is automatically the sum of the physical Hamiltonian (depending only on
Dirac's observables) and a gauge Hamiltonian (depending only on the gauge
variables and the 1st class constraints): in these cases there is a decoupling 
of the gauge degrees of freedom without the need to add gauge-fixings (i.e. 
without putting to zero the Abelianized gauge variables).

Therefore, given a system with constraints without a compact configuration 
space, one has to investigate whether there is any obstruction to the existence 
of global Shanmugadhasan canonical transformations, namely whether one can do 
a global canonical reduction. Possible obstructions can arise when the
configuration (and then also the phase) space and/or related mathematical 
structures like fiber bundles are not topologically trivial: usually they may 
be present when certain cohomological groups of the classical manifolds are not
trivial (these groups are at the basis of the possible existence of anomalies 
in the quantization of the system). A related problem in gauge theories is the
possible existence of Gribov-type ambiguities (nontrivial stability groups of
gauge transformations for gauge potentials and/or field strengths): since they
imply the nonexistence of global gauges, they destroy the possibility of a
global decoupling of physical and gauge degrees of freedom. These ambiguities
imply that both the configuration space and the phase space constraint manifold
are in general stratified manifolds (i.e. disjoint union of manifolds) possibly
with singularities. Moreover, for constrained systems defined in Minkowski
spacetime the constraint manifold is always a stratified manifold, because one
always assumes that the kinematical Poincar\'e group is globally implemented 
for isolated systems. This implies that the ten Poincar\'e generators must be 
finite (this is a first restriction on the boundary conditions of the fields
present in the isolated system to allow use of group theory) and that the
constraint manifold is the disjoint union of manifolds, each one of which 
contains all the system configurations belonging to the same type of Poincar\'e
orbit (spacelike orbits should be absent at the classical level not to have 
causality problems). The main stratum, dense in the constraint manifold, will 
be the one associated with nonzero spin (i.e. with nonzero Pauli-Lubanski
Casimir) timelike orbits. The existence of this Poincar\'e stratification 
raises the general question whether there could exist generalized Coulomb 
gauges with some kind of manifest covariance under Lorentz transformations and 
with some kind of universal breaking of manifest Lorentz covariance (unavoidable
when one eliminates all the gauge degrees of freedom).

Moreover, one would like to have of the results obtained in Minkowski spacetime
$M^4$ described in a form which can be extended to incorporate general 
relativity.

Given this general setting for constrained systems, a research program started 
trying to get a description only in terms of Dirac's observables and with an
explicit control on covariance of (to start with) the $SU(3)\times SU(2)\times 
U(1)$ standard model of elementary particles coupled to tetrad gravity (more 
natural than metric gravity for the coupling to fermion fields). See
Refs.\cite{re} for the genesis and the developments of this program, which is
well defined only for isolated systems [to recover theories with external 
fields one should consider special limits of some parameter of some subsystem].

In the next Sections a review of the results will be given.

\section{Noncovariant Generalized Coulomb Gauges.}

Dirac\cite{dira} found the Coulomb gauge physical Hamiltonian of the isolated
system formed by a fermion field plus the electromagnetic field [see the second
paper in Ref.\cite{lusa} for the case of Grassmann-valued fermion fields],
which contains the coupling of physical fermions with the radiation field and
the nonlocal Coulomb self-energy of the fermion field: $\int
d^3xd^3y\, ({\check \psi}^{\dagger}\check \psi )(\vec x,x^o){1\over {4\pi |\vec 
x-\vec y|}}({\check \psi}^{\dagger}\check \psi )(\vec y,x^o)$. The Dirac
observables are the transverse vector potential ${\vec A}_{\perp}(\vec x,x^o)$,
the transverse electric field ${\vec E}_{\perp}(\vec x,x^o)$ and physical
fermion fields dressed with a Coulomb cloud, $\check \psi (\vec x,x^o)=
e^{i\eta_{em}(\vec x,x^o)}\psi (\vec x,x^o)$, $\eta_{em}=-{1\over {\triangle}}
\vec \partial \cdot \vec A$. 

Extending this approach, the generalized Coulomb gauge of the following isolated
systems has been found [see Ref.\cite{re} c) for other systems like the Nambu 
string, relativistic two-body systems with action-at-a-distance interactions 
and for nonrelativistic Newton mechanics reformulated with 1st class 
constraints]:

a) Yang-Mills theory with Grassmann-valued fermion fields\cite{lusa} 
in the case of a trivial principal
bundle over a fixed-$x^o$ $R^3$ slice of Minkowski spacetime with suitable
Hamiltonian-oriented boundary conditions; this excludes monopole solutions and,
since $R^3$ is not compactified, one has only winding number and no instanton
number. After a discussion of the
Hamiltonian formulation of Yang-Mills theory, of its group of gauge
transformations and of the Gribov ambiguity, the theory has been studied in
suitable  weigthed Sobolev spaces where the Gribov ambiguity is absent.
The global Dirac observables are the transverse quantities ${\vec A}_{a\perp}
(\vec x,x^o)$, ${\vec E}_{a\perp}(\vec x,x^o)$ and fermion fields dressed
with Yang-Mills (gluonic) clouds.
The nonlocal and nonpolynomial (due to classical Wilson lines along flat 
geodesics) physical Hamiltonian has been obtained: it is nonlocal but without 
any kind of singularities, it has the correct Abelian limit if the structure 
constants are turned off, and it contains the explicit realization of the 
abstract Mitter-Viallet metric.

b) The Abelian and non-Abelian SU(2)
Higgs models with fermion fields\cite{lv1,lv2}, where the
symplectic decoupling is a refinement of the concept of unitary gauge.
There is an ambiguity in the solutions of the Gauss law constraints, which
reflects the existence of disjoint sectors of solutions of the Euler-Lagrange
equations of Higgs models. The physical Hamiltonian and Lagrangian of  the
Higgs phase have been found; the self-energy turns out to be local and
contains a local four-fermion interaction. 

c) The standard SU(3)xSU(2)xU(1) model of elementary particles\cite{lv3}
with Grassmann-valued fermion fields.
The final reduced Hamiltonian contains nonlocal self-energies for the
electromagnetic and color interactions, but ``local ones" for the weak 
interactions implying the nonperturbative emergence of 4-fermions interactions.
 To obtain a nonlocal self-energy with a Yukawa kernel for the
massive Z and $W^{\pm}$ bosons one has to reformulate the model on spacelike
hypersurfaces and make a modification of the Lagrangian.

\section{Wigner-Covariant Rest-Frame Instant Form.}

The next problem is how to covariantize these results. Again the starting point 
was given by Dirac\cite{dirac} with his reformulation of classical field theory 
on spacelike hypersurfaces foliating Minkowski spacetime. In this way one gets 
parametrized field theory with a covariant 3+1 splitting of flat spacetime and
already in a form suited to the coupling to general relativity in its ADM
canonical formulation (see also Ref.\cite{kuchar}
, where a theoretical study of this problem is done in curved spacetimes)
The price is that one has to add as new configuration variables  the points 
$z^{\mu}(\tau ,\vec \sigma )$ of the spacelike hypersurface $\Sigma_{\tau}$ 
[the only ones carrying Lorentz indices; the scalar parameter $\tau$ labels
the leaves of the foliation and $\vec \sigma$ are curvilinear coordinates on
$\Sigma_{\tau}$] and then to define the fields on
$\Sigma_{\tau}$ so that they know  the hypersurface $\Sigma_{\tau}$ of 
$\tau$-simultaneity [for a Klein-Gordon field $\phi (x)$ this new field is
$\tilde \phi (\tau ,\vec \sigma )=\phi (z(\tau ,\vec \sigma ))$]. Then,
besides a Lorentz-scalar form of the constraints of the given system, 
from the Lagrangian rewritten on the hypersurface [function of  $z^{\mu}$
through the induced metric $g_{\check A\check B}=z^{\mu}_{\check A}\eta_{\mu\nu}
z^{\nu}_{\check B}$, $z^{\mu}_{\check A}=\partial z^{\mu}/\partial \sigma
^{\check A}$, $\sigma^{\check A}=(\tau ,\sigma^{\check r})$] one gets
4 further first class constraints ${\cal H}_{\mu}(\tau ,\vec \sigma )
\approx 0$ implying the independence of the description from the choice of the 
spacelike hypersufaces. Being in special relativity, it is 
convenient to restrict ourselves to arbitrary spacelike hyperplanes $z^{\mu}
(\tau ,\vec \sigma )=x^{\mu}_s(\tau )+b^{\mu}_{\check r}(\tau ) \sigma^{\check 
r}$. Since they are described by only 10 variables [an origin $x^{\mu}_s(\tau 
)$ and 3 orthogonal spacelike unit vectors generating the fixed constant 
timelike unit normal to the hyperplane], we remain only with 10 first class
constraints determining the 10 variables conjugate to the hyperplane [they are
a 4-momentum $p^{\mu}_s$ and the 6 independent degrees of freedom hidden in a
spin tensor $S^{\mu\nu}_s$] in terms of the variables of the system.

If we now restrict ourselves to timelike ($p^2_s > 0$) 4-momenta, we can 
restrict the description to the so-called Wigner hyperplanes orthogonal to 
$p^{\mu}_s$ itself. To get this result, we must boost at rest all the 
variables with Lorentz indices by using the standard Wigner boost $L^{\mu}{}
_{\nu}(p_s,{\buildrel \circ \over p}_s)$ for timelike Poincar\'e orbits, and
then add the gauge-fixings $b^{\mu}_{\check r}(\tau )-L^{\mu}{}_{\check r}(p_s,
{\buildrel \circ \over p}_s)\approx 0$. Since these gauge-fixings depend on 
$p^{\mu}_s$, the final canonical variables, apart $p^{\mu}_s$ itself, are of 3
types: i) there is a non-covariant center-of-mass variable ${\tilde x}^{\mu}
(\tau )$ [the classical basis of the Newton-Wigner position operator]; ii) all
the 3-vector variables become Wigner spin 1 3-vectors [boosts in $M^4$ induce
Wigner rotations on them]; iii) all the other variables are Lorentz scalars. 
Only the 4 1st class constraints determining $p^{\mu}_s$ are left. One obtains 
in this way a new kind of instant form of the dynamics (see Ref.\cite{dira2}), 
the  Wigner-covariant 1-time rest-frame instant form\cite{lus1} with a 
universal breaking of Lorentz covariance. 
It is the special relativistic generalization of
the nonrelativistic separation of the center of mass from the relative motion
[$H={{ {\vec P}^2}\over {2M}}+H_{rel}$]. The role of the center of mass is 
taken by the Wigner hyperplane, identified by the point ${\tilde x}^{\mu}(\tau 
)$ and by its normal $p^{\mu}_s$. 
The 4 first class constraints can be put in the 
following form: i) the vanishing of the total (Wigner spin 1) 3-momentum of the
system $\vec p[system]\approx 0$ , saying that 
the Wigner hyperplane $\Sigma_W(\tau )$ is the intrinsic rest frame
[instead, ${\vec p}_s$ is left arbitrary, since it reflects the orientation of
the Wigner hyperplane with respect to arbitrary reference frames in $M^4$]; 
ii) $\pm \sqrt{p^2_s}-M[system]\approx 0$, saying that the
invariant mass M of the system replaces the nonrelativistic  Hamiltonian
$H_{rel}$ for the relative degrees of freedom, after the addition of the
gauge-fixing $T_s-\tau \approx 0$ [identifying the time parameter $\tau$ with 
the Lorentz scalar time of the center of mass in the rest frame; M generates the
evolution in this time]. When one is able, as in the case of N free particles
\cite{lus1},
to find the (Wigner spin 1) 3-vector $\vec \eta (\tau )$ conjugate to
$\vec p[system]$($\approx 0$), the gauge-fixing $\vec \eta \approx 0$ eliminates
the gauge variables describing the 3-dimensional 
intrinsic center of mass inside the
Wigner hyperplane [$\vec \eta \approx 0$ forces it to coincide with $x^{\mu}_s
(\tau )=z^{\mu}(\tau ,\vec \sigma =\vec \eta =0)$ and breaks the translation
invariance $\vec \sigma \mapsto \vec \sigma +\vec a$], so that we remain only 
with Newtonian-like degrees of freedom with rotational covariance: i) a 
3-coordinate (not Lorentz covariant) ${\vec z}_s=\sqrt{p_s^2}({\vec {\tilde x}}
_s-{{{\vec p}_s}\over {p_s^o}}{\tilde x}^o)$ and its conjugate momentum 
${\vec k}_s={\vec p}_s/\sqrt{p^2_s}$ 
for the absolute center of mass in $M^4$; ii) a set of relative
conjugate pairs of variables with Wigner covariance inside the Wigner hyperplane
.

The systems till now analyzed to get their rest-frame generalized
Coulomb gauges  are:

a) The system of N scalar particles with Grassmann electric charges
plus the electromagnetic field \cite{lus1}. The starting configuration 
variables are a 3-vector ${\vec 
\eta}_i(\tau )$ for each particle [$x^{\mu}_i(\tau )=z^{\mu}(\tau ,{\vec \eta}
_i(\tau ))$] and the electromagnetic gauge potentials 
$A_{\check A}(\tau ,\vec \sigma )={{\partial z^{\mu}(\tau ,\vec \sigma )}\over
{\partial \sigma^{\check A}}} A_{\mu}(z(\tau ,\vec \sigma ))$, 
which know implicitely the embedding of
$\Sigma_{\tau}$ into $M^4$. One has to choose the sign of the energy of each
particle, because there are not mass-shell constraints (like $p_i^2-m^2_i\approx
0$) among the constraints of this formulation, due to the fact that one has only
3 degrees of freedom for particle, determining the intersection of a timelike
trajectory and of the spacelike hypersurface $\Sigma_{\tau}$. The final Dirac's 
observables are: i) the transverse radiation field variables; ii) the particle
canonical variables ${\vec \eta}_i(\tau )$, ${\check {\vec \kappa}}_i(\tau )$,
dressed with a Coulomb cloud. The physical Hamiltonian contains the Coulomb 
potentials extracted from field theory and there is a regularization of the
Coulomb self-energies due to the Grassmann character of the electric charges
$Q_i$ [$Q^2_i=0$]. In Ref.\cite{lus2} there is the study of the 
Lienard-Wiechert potentials and of Abraham-Lorentz-Dirac equations in this
rest-frame Coulomb gauge and also scalar electrodynamics is reformulated in it.
Also the rest-frame 1-time relativistic statistical mechanics is developed
\cite{lus1}.

b) The system of N scalar particles with Grassmann-valued color charges plus 
the color SU(3) Yang-Mills field\cite{lus3}: 
it gives the pseudoclassical descrption of the
relativistic scalar-quark model, deduced from the classical QCD Lagrangian and 
with the color field present. The physical invariant mass of the system is
given in terms of the Dirac observables. From the reduced Hamilton equations  
the second order equations of motion both for the reduced transverse color 
field and the particles are extracted. Then, one studies  the N=2 
(meson) case. A special form of the requirement of having only color singlets, 
suited for a field-independent quark model, produces a ``pseudoclassical 
asymptotic freedom" and a regularization of the quark self-energy. With these
results one can covariantize the bosonic part of the standard model given in
Ref.\cite{lv3}.
 
c) It is in an advanced stage the description of Dirac and chiral fields and
of spinning particles on spacelike hypersurfaces\cite{dep}. After its
completion, the rest-frame form of the full standard $SU(3)\times SU(2)\times
U(1)$ model can be achieved.

Finally,  to eliminate the three 1st class constraints $\vec p[system]
\approx 0$ by finding their natural gauge-fixings, when fields are present,
one needs to find a rest-frame canonical basis of center-of-mass and relative
variables for fields (in analogy to particles). Such a basis has already been
found for a real Klein-Gordon field\cite{lon}. This kind of basis will allow,
after quantization, to find the  asymptotic states of the covariant
Tomonaga-Schwinger formulation of quantum field theory on spacelike
hypersurfaces: these states are needed for the theory of quantum bound states
[since Fock states do not constitute a Cauchy problem for the field equations,
because an in (or out) particle can be in the absolute future of another one due
to the tensor product nature of these asymptotic states, bound state equations
like the Bethe-Salpeter one have spurious solutions which are excitations in
relative energies, the variables conjugate to relative times (which are gauge
variables\cite{lus1})].

\section{Ultraviolet Cutoff.}

As said in Ref.\cite{lus2,lus3}, the quantization of these rest-frame
models has to overcome two problems. On the particle
side, the complication is the quantization of the square roots associated
with the relativistic kinetic energy terms. On the field side (all physical
Hamiltonian are nonlocal and, with the exception of the Abelian case,
nonpolynomial), the obstacle
is the absence (notwithstanding there is no  no-go theorem) of a complete
regularization and renormalization procedure of electrodynamics (to start with) 
in the Coulomb gauge: see Ref.\cite{cou} (and its bibliography)
for the existing results for QED. 

However, as shown in Refs.\cite{lus1,lusa}
[see their bibliography for the relevent references referring to all the 
quantities introduced in this Section], the rest-frame instant 
form of dynamics automatically gives a physical ultraviolet cutoff in the 
spirit of Dirac and Yukawa: it is the M$\o$ller radius\cite{mol} 
$\rho =\sqrt{-W^2}c/P^2=|\vec S|c/\sqrt{P^2}$ ($W^2=-P^2{\vec 
S}^2$ is the Pauli-Lubanski Casimir), namely the classical intrinsic radius of 
the worldtube, around the covariant noncanonical Fokker-Price center of
inertia $Y^{\mu}$, 
inside which the noncovariance of the canonical center of mass ${\tilde
x}^{\mu}$ is concentrated. At the quantum level $\rho$ becomes the Compton 
wavelength of the isolated system multiplied its spin eigenvalue $\sqrt{s(s+1)}$
, $\rho \mapsto \hat \rho = \sqrt{s(s+1)} \hbar /M=\sqrt{s(s+1)} \lambda_M$ 
with $M=\sqrt{P^2}$ the invariant mass and $\lambda_M=\hbar /M$ its Compton
wavelength. Therefore, the criticism to classical relativistic physics, based
on quantum pair production, concerns the testing of distances where, due to the
Lorentz signature of spacetime, one has intrinsic classical covariance problems:
it is impossible to localize the canonical center of mass ${\tilde x}^{\mu}$
(also named Pryce center of mass and having the same covariance of the 
Newton-Wigner position operator) in a frame independent way.

Since $\rho$ describes a nontestable classical short distance
region [there is a conceptual connection with the aspect of Mach's principle
according to which only relative motions are measurable], it sounds reasonable 
\cite{lus3} that for a confined system of effective radius $r_o=
1/\Lambda_{QCD}$ (the fundamental scale of QCD) one has $\rho \leq
r^2_oM =M/\Lambda^2_{QCD}$ [this ensures the mass-spin relation
$|\vec S| = \alpha_s^{'} M^2+\alpha_o$ of phenomenological Regge
trajectories]. Let us note that in string theory\cite{ven}
the relevant dimensional quantity is the
tension $T_s=1/2\pi \alpha^{'}_s$ (the energy per unit length), which, at the 
quantum level, determines a minimal length $L_s=\sqrt{\hbar /T_s} =\sqrt{2\pi
\hbar \alpha^{'}_s}{\buildrel {\hbar =1} \over =}\, \sqrt{2\pi \alpha^{'}_s}$ 
[for a classical string one has $|\vec S| \leq \alpha_s^{'} M^2$; a QCD
string has $2\pi \alpha^{'}_s \leq r^2_o=\Lambda^{-2}_{QCD}$].

Let us remember \cite{lus1}
that $\rho$ is also a remnant in flat Minkowski spacetime of 
the energy conditions of general relativity: since the M$\o$ller
noncanonical, noncovariant center of energy has its noncovariance localized
inside the same worldtube with radius $\rho$ (it was discovered in this way)
\cite{mol}, it turns out that an extended relativistic system with the
material radius smaller of its intrinsic radius $\rho$ has: i) the peripheral
rotation velocity can exceed the velocity of light; ii) its classical energy
density cannot be positive definite everywhere in every frame. Now, the real
relevant point is that this ultraviolet cutoff determined by
$\rho$ exists also in Einstein's
general relativity (which is not power counting renormalizable) in the case of
asymptotically flat spacetimes, taking into account the Poincar\'e Casimirs of
its asymptotic ADM Poincar\'e charges (when supertranslations are eliminated 
with suitable boundary conditions; let us remark that Einstein and Wheeler
use closed universes because they don't want to introduce boundary conditions,
but in this way they loose Poincar\'e charges and the possibility to make 
contact  with particle physics).

By comparison, in string cosmology\cite{ven}, at 
the quantum level, the string tension $T_{cs}=1/2\pi \alpha^{'}_{cs}=
L^2_{cs}/\hbar$ gives
rise to a minimal length $L_{cs}{\buildrel {\hbar =1} \over =}\, \sqrt{2\pi 
\alpha^{'}_{cs}} \geq L_P$ [$L_P=1.6\, 10^{-33} cm$ is 
the Planck length] and is determined by the vacuum 
expectation value of the background metric of the vacuum (if the ground state is
flat Minkowski spacetime), while the grand unified coupling constant $\alpha
_{GUT}$ (replacing $\alpha_s$ of QCD) is determined by the vacuum expectation 
value of the background dilaton field. This minimal length $L_{cs}\geq
L_P$ (suppressing the gravitational corrections) could be a lower bound 
for the M$\o$ller radius of an asymptotically flat universe. The upper bound 
on $\rho$ (namely a physical infrared cutoff) could be the Hubble distance 
$cH_o^{-1}\approx 10^{28} cm$ considered as an effective radius of the universe.
Therefore, it seems reasonable that our physical ultraviolet cutoff $\rho$
is meaningful in the range $L_P\leq L_{cs} < \rho < cH_o^{-1}$.

Moreover, the extended Heisenberg relations  of string theory\cite{ven}, i.e.
$\triangle x ={{\hbar}\over {\triangle p}}+{{\triangle p}\over {T_{cs}}}=
{{\hbar}\over {\triangle p}}+{{\hbar \triangle p}\over {L^2_{cs}}}$ implying the
lower bound $\triangle x > L_{cs}=\sqrt{\hbar /T_{cs}}$ due to the $y+1/y$
structure,
have a counterpart in the quantization of the M$\o$ller radius\cite{lus1}:
if we ask that, also at the quantum level, one cannot test the inside of the 
worldtube, we must ask $\triangle x > \hat \rho$ which is the lower bound
implied by the modified uncertainty relation $\triangle x ={{\hbar}\over 
{\triangle p}}+{{\hbar \triangle p}\over {{\hat \rho}^2}}$. This would imply 
that the center-of-mass canonical noncovariant (Pryce) 3-coordinate 
$\vec z=\sqrt{P^2}({\vec {\tilde x}}-{{\vec P}\over {P^o}}{\tilde x}^o)$ 
\cite{lus1} cannot become a
self-adjoint operator. See Hegerfeldt's theorems (quoted in 
Refs.\cite{lusa,lus1}) and his interpretation 
pointing at the impossibility of a good localization of relativistic particles
(experimentally one determines only a worldtube in spacetime emerging from the 
interaction region). Since the eigenfunctions of the canonical center-of-mass
operator are playing the role of the wave function of the universe, one could 
also say that the center-of-mass variable has not to be quantized, because it
lies on the classical macroscopic side of Copenhagen's interpretation and,
moreover, because, in the spirit of Mach's principle that only relative 
motions can be observed, no one can observe it. On the other hand, if one 
rejects the canonical noncovariant center of mass in favor of the covariant
noncanonical Fokker-Pryce center of inertia $Y^{\mu}$, $\{ Y^{\mu},Y^{\nu} \}
\not= 0$, one could invoke the philosophy of quantum groups to quantize 
$Y^{\mu}$ to get some kind of quantum plane for the center-of-mass 
description.

\section{Tetrad Gravity.}

The next step of the program is the search of Dirac's observables for classical
tetrad gravity in globally hyperbolic asymptotically flat spacetimes $M^4=
\Sigma \times R$ with $\Sigma$ diffeomorphic to $R^3$, so to have the 
asymptotic Poincar\'e charges and the same ultraviolet cutoff $\rho$ as for
the other interactions.

In Ref.\cite{rus}  there is a new formulation of tetrad gravity avoiding the
use of Schwinger's time gauge condition and, with the technology developed for
Yang-Mills theory\cite{lusa}, 13 of its 14 ${}$ 1st class constraints have been 
Abelianized [the Abelianization of the 6 constraints generating 
space-diffeomorphisms and Lorentz rotations has been done in 3-orthogonal 
coordinates on $\Sigma$ so that the 3-metric is diagonal]. The last constraint
(the superHamiltonian one) becomes an integral equation for the momentum
conjugate to the conformal factor of the 3-metric. See Ref.\cite{re} c) for an
expanded summary of the results and of the still open problems.

Further problems are how to deparametrize the theory\cite{isha}, so to
reexpress it in the form of parametrized field theories on spacelike
hypersurfaces in Minkowski spacetime. This is an extremely important point,
because, if we add N scalar particles to tetrad gravity (whose reduction to
Dirac's observables should define the N-body problem in general relativity),
the deparametrization should be the bridge to the previously quoted theory on
spacelike hypersurfaces in Minkowski spacetime\cite{lus1,lus2,lus3} in the 
limit of
zero curvature. A new formulation of the N-body problem would be relevant to
try to understand the energy balance in the emission of gravitational waves
from systems like binaries. If it will be possible to find the Dirac
observables for the particles, one will understand how to extract from the
field theory the covariantization of Newton potential [one expects one scalar
and one vector (gravitomagnetism) potential] and a mayor problem will be how to 
face the expected singularities of the mass-self-energies.

Finally one should couple tetrad gravity to the
electromagnetic field, to fermion fields and then to the standard model,
trying to make to reduction to Dirac's observables in all these cases.


\begin{references}
\bibitem{dirac}P.A.M.Dirac, Can.J.Math. {\bf 2}, 129 (1950); "Lectures on 
Quantum Mechanics", Belfer Graduate School of Science, Monographs Series 
(Yeshiva University, New York, N.Y., 1964).
\bibitem{sha}S.Shanmugadhasan, J.Math.Phys. {\bf 14}, 677 (1973).
L.Lusanna, Int.J.Mod.Phys. {\bf A8}, 4193 (1993).
M.Chaichian, D.Louis Martinez and L.Lusanna, Ann.Phys.(N.Y.){\bf 232}, 40 
(1994). L.Lusanna, 
Phys.Rep. {\bf 185}, 1 (1990); Riv. Nuovo Cimento {\bf 14}, n.3, 1 (1991);
J.Math.Phys. {\bf 31}, 2126 (1990); J.Math.Phys. 
{\bf 31}, 428 (1990).
\bibitem{re}L.Lusanna, a) "Classical Observables of Gauge Theories from the
Multitemporal Approach", talk given at the Conference 'Mathematical
Aspects of Classical Field Theory', Seattle 1991, in Contemporary
Mathematics {\bf 132}, 531 (1992);
b) ``Hamiltonian Constraints and Dirac's Observables: from
Relativistic Particles towards Field Theory and General Relativity",
talk at the Workshop ``Geometry of Constrained Dynamical Systems",
Newton Institute, Cambridge, 1994, (J.M.Charap, Ed.), Cambridge Univ.Press,
Cambridge, 1995.
c) ``Solving Gauss' Laws and Searching
Dirac Observables for the Four Interactions", talk at the ``Second Conf. on
Constrained Dynamics and Quantum Gravity", S.Margherita Ligure 1996 
(HEP-TH/9702114).
\bibitem{dira}P.A.M.Dirac, Can.J.Phys. {\bf 33}, 650 (1955).
\bibitem{lusa}L.Lusanna, Int.J.Mod.Phys. {\bf A10}, 3531 and 3675 (1995).
\bibitem{lv1}L.Lusanna and P.Valtancoli, ``Dirac's Observables for the Higgs
model: I) the Abelian Case", to appear in Int.J.Mod.Phys. A (HEP-TH/9606078).
\bibitem{lv2}L.Lusanna and P.Valtancoli, ``Dirac's Observables for the Higgs
model: II) the non-Abelian SU(2) Case", to appear in Int.J.Mod.Phys. A
(HEP-TH/9606079).
\bibitem{lv3}L.Lusanna and P.Valtancoli, ``Dirac's Observables for the 
SU(3)xSU(2)xU(1) Standard Model", Firenze Univ.preprint, May 1997.
\bibitem{kuchar}K.Kuchar, J.Math.Phys. {\bf 17}, 777, 792, 801 (1976); {\bf 18},
1589 (1977).
\bibitem{dira2}P.A.M.Dirac, {\it Rev.Mod.Phys.} {\bf 21} (1949) 392.
\bibitem{lus1}L.Lusanna, Int.J.Mod.Phys. {\bf A12}, 645 (1997).
\bibitem{lus2}D.Alba and L.Lusanna, ``The Lienard-Wiechert Potential of Charged
Scalar Particles and their Relation to Scalar Electrodynamics in the Rest-Frame
Instant Form", Firenze Univ.preprint, May 1997.
\bibitem{lus3}D.Alba and L.Lusanna, ``The Classical Relativistic Quark Model in
the Rest-Frame Wigner-Covariant Coulomb Gauge", Firenze Univ.preprint,May 1997.
\bibitem{dep}F.Bigazzi, R.DePietri and L.Lusanna, in preparation.
\bibitem{lon}G.Longhi and M.Materassi, ``Collective and Relative Variables for
a Classical Relativistic Field", in preparation.                                
\bibitem{cou}
G.Leibbrandt, ``Non-Covariant Gauges", ch.9 (World Scientific, Singapore, 1994).
\bibitem{mol}C.M$\o$ller, Ann.Inst.H.Poincar\'e {\bf 11}, 251 (1949); ``The 
Theory of Relativity" (Oxford Univ.Press, Oxford, 1957).
\bibitem{ven}G.Veneziano, ``Quantum Strings and the Constants of Nature", in
``The Challenging Questions", ed.A.Zichichi, the Subnuclear Series n.27
(Plenum Press, New York, 1990). 
\bibitem{rus}L.Lusanna and S.Russo, ``Dirac's Observables for Tetrad Gravity",
in preparation.
\bibitem{isha}C.J.Isham and K.Kuchar, Ann.Phys.(N.Y.) {\bf 164}, 288 and 316
(1984). K.Kuchar, Found.Phys. {\bf 16}, 193 (1986).


\end{references}
\end{document}